# Untangling the Global Coronal Magnetic Field with Multiwavelength Observations

S. E. Gibson, A. Malanushenko, G. de Toma, S. Tomczyk (NCAR/HAO); K. Reeves (Smithsonian CfA); H. Tian, Z. Yang (Peking Univ.); B. Chen, G. Fleishman, D. Gary, G. Nita (NJIT); V. M. Pillet (NSO); S. White (AFRL); U. Bąk-Stęślicka (Wroclaw Univ.); K. Dalmasse (IRAP); T. Kucera (NASA/GSFC); L. A. Rachmeler (NOAA); N. E. Raouafi (JHU/APL); J. Zhao (Purple Mountain Observatory)

Magnetism defines the complex and dynamic solar corona. Coronal mass ejections (CMEs) are thought to be caused by stresses, twists, and tangles in coronal magnetic fields that build up energy and ultimately erupt, hurling plasma into interplanetary space. Even the ever-present solar wind possesses a three-dimensional morphology shaped by the global coronal magnetic field, forming geoeffective corotating interaction regions. CME evolution and the structure of the solar wind depend intimately on the coronal magnetic field, so comprehensive observations of the global magnetothermal atmosphere are crucial both for scientific progress and space weather predictions. Although some advances have been made in measuring coronal magnetic fields locally, synoptic measurements of the global coronal magnetic field are not yet available.

For decades we have observed the magnetic field at the solar surface (photosphere) with ever-increasing spatial and temporal resolution. From these observations we have learned much about how magnetic flux emerges and evolves over multiple time scales. We know that magnetic forces in the solar atmosphere—especially in the corona—dominate over plasma forces. However, until very recently our knowledge of coronal magnetic fields was limited to what we could infer from solar surface measurements and from coronal plasma morphology.

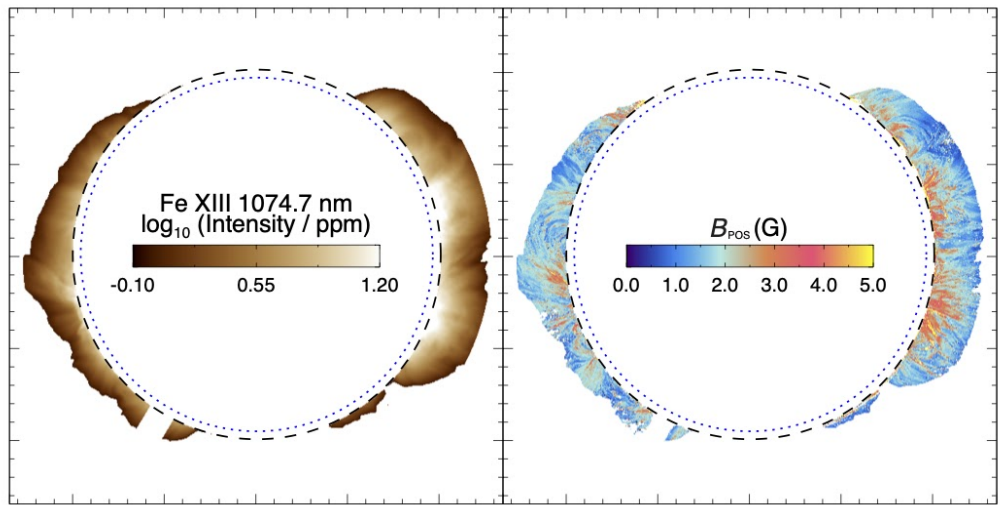

***Fig. 1.*** *A coronal image from the Mauna Loa Solar Observatory's Coronal Multichannel Polarimeter (CoMP) (left) and the corresponding plane-of-sky magnetic field map (right). Yang et al. 2020a, Science.*

This is no longer the case. Just this year coronal magnetism has been mapped as never before, using visible/infrared (VIR) coronal seismology (*Yang et al. 2020a, 2020b;* Figure 1) and microwave interferometry (*Fleishman et al. 2020*). The 4-meter Daniel K. Inouye Solar Telescope (DKIST, *Rimmele et al. 2020*) has sufficient light-gathering capability to observe the line-of-sight component of the coronal magnetic field in VIR (*Fehlmann et al. 2016*), as does the proposed full-Sun Coronal Solar Magnetism Observatory (COSMO; *Tomczyk et al. 2016*) which would enable daily maps of the global coronal magnetic fields in three dimensions. The Expanded Owens Valley Solar Array (EOVSA) is enabling unprecedented measurements of coronal magnetism during solar flares (*Chen et al. 2020*), and the proposed Frequency Agile

Solar Radiotelescope (FASR; *Bastian et al. 2003*) would provide daily global mapping of the magnetized corona both on-disk and off-limb. The capabilities at short wavelengths are largely unexplored, but the potential of UV wavelengths for diagnosing magnetic field strength using ultraviolet (UV) spectropolarimetry has been demonstrated through clever use of the SOHO spacecraft roll (*Raouafi et al. 2002*), and forward modeling has indicated that a space-based UV spectropolarimeter would be a powerful complementary constraint on the three-dimensional off-limb coronal magnetic field (*Zhao et al. 2019*). Finally, a magnetic-field-induced atomic transition has recently been identified from EUV spectra, demonstrating its potential for diagnosing the coronal magnetic field (*Li et al. 2015, Landi et al. 2020; Ran et al. 2020*).

| Process | Physical-state dependency | Observation | Magnetic quantity probed |
|---|---|---|---|
| Thomson scattering | electron density | White-light pB, TB | Plasma structured by field (e.g. closed vs. open field boundaries, flux surfaces) |
| Collisional excitation | electron density, temperature | IR/Visible/EUV/SXR emission | Plasma structured by field (incl. loops, closed/open boundaries, flux surfaces) |
| Continuum absorption | chromospheric population density, electron density, temperature | EUV absorption features | Can indicate magnetic geometry suitable for prominence formation |
| Resonance scattering; polarization | electron density, temperature, vector magnetic field | Visible/IR spectra | $B_{los}$ from Stokes V; Magnetic field direction from Stokes Q, U |
| Doppler shift | electron density, temperature, velocity | Visible/IR spectra | $B_{pos}$ and field line direction from waves; flux surfaces from bulk flows |
| Thermal bremsstrahllung | electron density, temperature, vector magnetic field | Radio emission (intensity and circular polarization) as a function of frequency | $B_{los}$ from Stokes V |
| Gyroresonance | electron density, temperature, vector magnetic field | Radio emission (intensity and circular polarization) as a function of frequency | Surfaces of constant magnetic field strength at each frequency |
| Faraday rotation | electron density, temperature, vector magnetic field | Rotation of plane of polarization | $B_{los}$ from rotation measure |

**Table 1.** A subset of the magnetically-sensitive physical processes operating in the solar corona, highlighting dependency on attributes of the physical state, the observations sensitive to these processes, and diagnostic sensitivity of the observables to the 3D coronal magnetic field. From *Gibson et al. (2016)*.

**Multiwavelength observations provide crucial building blocks needed to construct a comprehensive picture of the coronal magnetic field.** VIR observations probe magnetic fields at different temperatures and at higher heights in the coronal atmosphere than radio observations can, while radio observations, which do not require an occulting disk, provide information in the lower corona and chromosphere. The linear polarization of coronal VIR emission lines in the saturated regime of the Hanle effect provides information about the direction of the coronal magnetic field as projected on the plane of the sky (POS) and, consequently, about the coronal topology (*Bąk-Stęślicka et al. 2013; Rachmeler et al. 2014*). Complentarily, the linear polarization of UV lines in the unsaturated Hanle regime are dependent on both the strength of the magnetic field and its component along the line of sight (LOS) (*Fineschi et al.,1999; Zhao et al. 2019*). LOS magnetic field strength, integrated through the

optically-thin corona, may also be obtained from VIR emission of magnetic dipole lines (*Lin et al. 2004; Fan et al. 2018*), or alternatively from free-free emission in radio (when combined with a density model; *Gelfreikh 2004*). Magneto-acoustic waves (at various wavelengths) also provide a diagnostic of POS magnetic field strength when combined with observed or modeled density (*Tomczyk et al. 2007; Yang et al. 2020a, 2020b*).

Pulling together all of these measurements to infer the three-dimensional magnetic field is a global optimization problem: given magnetically-sensitive coronal observations, determine the magnetic field distribution that generates them. Complexities arise because different observations have different dependencies on the physical state and thus provide different diagnostics of the coronal magnetic field (see **Table 1**). Community tools like the FORWARD and GX_Simulator software packages, which synthesize coronal observables based on model/simulation input (Nita et al. 2015, 2018; Gibson et al. 2016), help pull these strands together and may be used by inversion frameworks along with observations to solve for an optimized coronal magnetic field (e.g., Kramar et al., 2006, 2013, 2016; Dalmasse et al. 2019).

***We conclude that a key goal for 2050 should be comprehensive, ongoing 3D synoptic maps of the global coronal magnetic field.*** This will require the construction of new telescopes, ground and space-based, to obtain complementary, multiwavelength observations sensitive to the coronal magnetic field. It will also require development of inversion frameworks capable of incorporating multi-wavelength data, and forward analysis tools and simulation testbeds to prioritize and establish observational requirements on the proposed telescopes.

**References.**
Bąk-Stęślicka, U. et al., *ApJL*, 770, L28, 2013
Bastian, T. S., *ASR*, 32, 12, 2705, 2003
Chen, B. et al., *Nature Astron.*, DOI:10.1038/s41550-020-1147-7/2020
Fleishman et al., *Science*, 367, 6475, 278, 2020
Gelfreikh, G. B., in *Solar and Space Weather Radiophysics*, (Dordrecht), 115, 2004
Gibson et al., *FrASS,* 3, 8, 2016
Gibson et al., *ApJL*, 840, L13, 2017
Kramar, M., Inhester, B., Solanki, S. K., *A&A*, 456, 665, 2006
Kramar, M., Inhester, B., Lin, H., Davila, J., *ApJ*, 775, 25, 2013
Kramar, M., Lin, H., Tomczyk, S., *ApJL*, 819, 36. 2016
Landi, E., et al. submitted to *ApJ*, 2020
Li, W., et al., *ApJ*, 807, 69, 2015
Lin, H., Kuhn, J. R., and Coulter, R., *ApJL*, 613, L177, 2004
Nita et al., *ApJ*, 799 (2), 236, 2015
Nita et al., *ApJ,* 853 (1), 66, 2018
Rachmeler, L. et al., *ApJL*, 787, L3, 2014
Ran et al., *ApJL*, 898, 34, 2020
Raouafi, N.-E., Sahal-Bréchot, S., Lemaire, P., *A&A,* 396, 1019, 2002
Rimmele, T. et al., Solar Physics, in press, 2020
Tomczyk, S. et al., *Science*, 317,1192, 2017
Tomczyk, S. et al., *JGRA,* 121, 7470, 2017
Yang, Z. et al., *Science*, 369, 694, 2020
Yang, Z. et al., *Sci China Tech Sci,* DOI: https://doi.org/10.1007/s11431-020-1706-9, 2020
Zhao, J. et al., *ApJ*, 883, 55, 2019